\documentclass[onecolumn,showpacs,amsmath,amssymb,prd,nofootinbib]{revtex4-2}
\usepackage{graphicx}
\usepackage{epsfig}
\usepackage{enumerate}

\def\beq{\begin{equation}}
\def\eeqno#1{\label{#1}\end{equation}}

\def\rarrow{\rightarrow }

\def\dleft{\rlap{{\it D}}\raise 8pt
\hbox{$\scriptscriptstyle\Leftarrow$}}
\def\dright{\rlap{{\it
D}}\raise 8pt\hbox{$\scriptscriptstyle\Rightarrow$}}

\def\cmss{{\rm cm~s^{-2}}}

\def\az{a_{0}}
\def\azs{a_{0}^2}

\def\l0{\ell_{0}}

\def\rar{\rightarrow}
\def\s{\sigma}

\def\a{\alpha}
\def\b{\beta}
\def\c{\gamma}
\def\l{\lambda}

\def\f{\phi}
\def\t{\theta}

\def\k{\kappa}
\def\r{\rho}
\def\m{\mu}
\def\n{\nu}
\def\z{\zeta}

\def\av#1{\langle#1\rangle}

\def\bo{\bar\omega}

\def\Ct{\textbf{C}}
\def\Qt{\textbf{Q}}

\def\Z{\mathcal{Z}}

\def\o{\omega}

\def\d{\delta}

\def\a{\alpha}
\def\xlimin{{x\rarrow\infty \atop{\raise 1pt\hbox to 30pt
{\rightarrowfill}}}}
\def\limlim#1#2{{#1\rarrow #2 \atop{\raise 1pt\hbox to 30pt
{\rightarrowfill}}}}
\def\eps{\epsilon}

\def\ve{{\bf e}}

\def\grad{\vec\nabla}
\def\div{\vec \nabla\cdot}

\def\gfs{(\grad\phi)^2}
\def\gfss{(\gf\^*)^2}

\def\pd#1#2{\frac{\partial#1}{\partial#2}}

\def\U{\mathcal{U}}
\def\V{\mathcal{V}}
\def\W{\mathcal{W}}
\def\Y{\mathcal{Y}}

\def\M{\mathcal{M}}
\def\m{\mu}
\def\a{\alpha}
\def\b{\beta}
\def\C{\Gamma}

\def\n{\nu}
\def\Up{\Upsilon}
\def\ten#1#2{\^{#1}\_{#2}}

\def\cd#1{{}_{;#1}}
\def\emn{\eta_{\m\n}}
\def\fh{\hat\f}

\def\fs{\f\^*}

\def\rh{\hat\rho}

\def\_#1{_{\scriptscriptstyle #1}}
\def\^#1{^{\scriptscriptstyle #1}}
\def\baz{\bar a_0}

\def\emn{\eta\_{\m\n}}

\def\gmn{g\_{\m\n}}
\def\Gmn{g\^{\mu \nu}}
\def\rg{\tilde g}
\def\rgt{{\bf\rg}}
\def\rgmn{\rg\_{\m\n}}
\def\rGmn{\rg\^{\m\n}}

\def\Gab{g\^{\alpha\beta}}

\def\hGmn{\hat g\^{\m\n}}
\def\hGab{\hat g\^{\a\b}}
\def\hgmn{\hat g\_{\m\n}}

\def\hmn{h\_{\m\n}}

\def\C{\Gamma}
\def\der#1{\_{,#1}}

\def\lM{\ell\_M}

\def\T{\mathcal{T}}
\def\Tmn{\T\_{\m\n}}
\def\Th{\hat{\mathcal{T}}}
\def\hTmn{\Th\_{\m\n}}
\def\oot{\frac{1}{2}}
\def\gt{\textbf{g}}
\def\hgt{\hat\gt}

\def\grp{\breve{g}}
\def\drp{\breve{\delta}}

\def \hij {h\_{ij}}

\def\hsij{h\^*\_{ij}}
\def\hs{h\^*}
\def \dij {\d\_{ij}}
\def\fpg{4\pi G}
\def\epg{8\pi G}
\def\spg{16\pi G}

\def\gfss{(\nabla\fs)\^2}
\def\vrf{\varphi}
\def\vrfzs{\vrf\der{0}\^2}
\def\gvrfs{(\grad\varphi)^2}
\begin{document}
\title{Broader view of bimetric MOND}

\author{Mordehai Milgrom}
\affiliation{Department of Particle Physics and Astrophysics, Weizmann Institute}

\begin{abstract}
All existing treatments of bimetric MOND (BIMOND) -- a class of relativistic versions of MOND -- have dealt with a rather restricted subclass: The Lagrangian of the interaction between the gravitational degrees of freedom -- the two metrics -- is a function of a certain {\it single} scalar argument built from the difference in connections of the two metrics. I show that the scope of BIMOND is much richer: The two metrics can couple through several scalars to give theories that all have a ``good''  nonrelativistic (NR) limit -- one that accounts correctly, {\it \`{a}-la} MOND, for the dynamics of galactic systems, {\it including gravitational lensing}. This extended-BIMOND framework exhibits a qualitative departure from the way we think of MOND at present, as encapsulated, in all its aspects, by one ``interpolating function''  of one acceleration variable.
After deriving the general field equations, I pinpoint the subclass of theories that satisfy the pivotal requirement of a good NR limit. These involve three independent, quadratic scalar variables. In the NR limit these scalars all reduce to the same acceleration scalar, and the NR theory then does hinge on one function of a {\it single} acceleration variable -- representing the NR MOND interpolating function, whose form is largely dictated by the observed NR galactic dynamics. However, these scalars take different values, and behave differently, in different relativistic contexts. So, the full richness of the multivariable Lagrangian, as it enters cosmology, for example, or gravitational-wave dynamics, is hardly informed by what we learn of MOND from observations of galactic dynamics. In this paper, I present the formalism, with some generic examples. I also consider some cosmological solutions where the two metrics are small departures from one Friedman-Lemaitre-Robertson-Walker metric. This may offer a framework for describing cosmology within the extended BIMOND.

\end{abstract}
\maketitle

\section{Introduction}
The  MOND paradigm \cite{milgrom83} extends Newtonian dynamics and general relativity, so as to account for the dynamics of galactic systems, and of the Universe at large, without ``dark matter.'' MOND proposes that standard dynamics break down in the limit where accelerations in a system are around or below a certain acceleration $\az$ -- the MOND acceleration constant. In the ``deep-MOND''  limit -- much below $\az$ -- MOND posits that dynamics become spacetime scale invariant, at least in the nonrelativistic (NR) limit, and for systems whose dynamics is governed by gravity \cite{milgrom09}. In the high-acceleration limit, MOND posits a rapid return to standard dynamics (a ``correspondence principle'').
Reviews of MOND can be found in Refs. \cite{fm12,milgrom14,milgrom20a,mcgaugh20,merritt20,bz22}.
\par
Already from the above basic axioms, a large number of ``MOND laws''  of galactic dynamics follow \cite{milgrom14a}. As discussed in detail in the above reviews, these predictions have been amply tested by observations of galactic systems.
The MOND constant appears in various roles in the MOND predictions, and so its value could be deduced from several independent observations, all giving, consistently, a value of $\az\approx 1.2\times 10\^{-8}\cmss$.
\par
Quite interestingly (e.g., Refs. \cite{milgrom83,milgrom20}), this value of $\az$ may be of cosmological significance, since
\beq \baz \equiv 2\pi \az\approx a\_H(0)\equiv cH_0\approx a\_\Lambda\equiv c^2/\ell\_{\Lambda}, \eeqno{coinc}
where $H_0$ is the present-day value of the cosmological expansion rate, and  $\ell\_{\Lambda}=(\Lambda/3)^{-1/2}$ is the radius associated with $\Lambda$ -- the observed equivalent of a cosmological constant. The ``MOND length,'' $ \lM\equiv c^2/\az$, is thus of the order of the present characteristic size of the observable universe, $\ell\_U$.
This numerical ``coincidence,'' if fundamental, may have far-reaching ramifications for MOND, and for gravity in general. It would also be exciting if this can be shown to follow from some fundamental MOND theory, as discussed in Ref. \cite{milgrom20}.
\par
At present, we work with two NR Lagrangian formulations of MOND that embody the above basic tenets, one dubbed AQUAL (for ``aquadratic Lagrangian'') \cite{bm84}, the other, QUMOND (for ``quasilinear MOND'') \cite{milgrom10}. These have been applied extensively to study (NR) dynamical processes in galactic systems, such as the formation and interactions of galaxies, dynamical friction, etc. (e.g., Refs. \cite{brada00,tiret08,candlish15,thomas17,bilek17,banik22,banik22a}).
\par
The dynamics of galactic systems is highly NR, and accounting for them within present accuracy does not require a relativistic theory: The highest mean-field gravitational potentials encountered in such systems (and the squared velocities virially related to them) are $\lesssim 10^{-5}c^2$.
However, it has been clear from the outset that we must construct relativistic theories whose NR limit complies with MOND's basic tenets.
General relativity (GR) reduces to Newtonian dynamics in the NR limit; so if MOND is to extend the latter in the NR regime, there must be a relativistic theory that extends GR and that describes MOND phenomenology in this limit. To boot, there are central astrophysical phenomena that do require a relativistic description.
\par
One of these concerns gravitational lensing by galactic systems and large-scale structure. We are then still in the weak-field limit of gravity itself -- the gravitational field is produced by slow moving bodies, and the potential is $\ll c^2$. But we are dealing with photon trajectories, and these probe the theory beyond the NR limit. In a single-metric theory, such as GR, there is a coordinate frame where
the metric describing the gravitational field of a NR system, such as a galaxy, can be brought to the form
\beq \gmn\approx\emn-2c^{-2}{\rm diag}(\f,\psi,\psi,\psi),   \eeqno{jamujam}
with $\f,~\psi\ll c^2$. The potential $\f$ alone governs the motion of slow-moving bodies, such as that of galactic constituents within a galactic system. But the null trajectories of massless particles (photons) depend also on $\psi$. In GR, $\psi=\f$, but this is not necessarily so in other relativistic theories. We can tell to what extent this equality holds in nature, by comparing the observed motion of slow particles -- e.g., of stars or gas rotating around a galaxy -- with gravitational lensing by the same galaxy, and see if they are governed by a single potential.
\par
If we want such a comparison to pertain to the MOND-vs-dark-matter issue, we need to make it in low-acceleration systems -- i.e., where so-called ``mass anomalies''  prevail according to MOND -- because MOND is posited to coincide very nearly with GR in the high-acceleration regime.
\par
Comparison of lensing with slow-moving test masses in galaxy clusters (e.g., Ref. \cite{hoekstra15}), and of lensing at the very outskirts of galaxies of all types, compared with their rotation curves (e.g., Refs. \cite{milgrom13,brouwer21,mcgaugh22}) is consistent with this single-potential NR geometry,\footnote{Although, the accuracy is not very high, and there is still room for some departure from this geometry.} as one finds in these cases $\psi\approx \f$. Given this, construction of relativistic MOND formulations has concentrated on theories whose NR dynamics is governed by a single potential $\f$.
In such MOND theories, the NR potential, $\f$, is then not determined by the Poisson equation, as in GR, but by some other, MOND, equation, such as those of AQUAL or QUMOND, mentioned above.
\par
Beyond weak-gravity systems -- such as galaxies and galactic systems -- what strong-gravity phenomena ($\f\not\ll c^2$) fall in the MOND regime?
Because of the above coincidence (\ref{coinc}), a system of size $R$ and characteristic potential $\f$ is both in the MOND regime -- i.e., having $\f/R\lesssim \az\sim c^2/\ell\_U$ -- and in the strong-gravity regime -- having $\f\sim c^2$ -- must have $R\gtrsim \ell\_U$, namely, a size comparable or larger than that of the observable Universe. Thus, the only strong-gravity, MOND system is the Universe at large and its cosmology. In ``cosmology''  we also include such aspects as structure formation and the cosmic microwave background, which are strictly speaking not strong field, but do require modified dynamics if we want to avoid dark components. On the other hand, strong-field phenomena, such as the formation and mergers of black holes, and the emission of gravitational waves in these processes, are strong-field phenomena, but occur at accelerations much higher than $\az$, and can hardly be affected by MOND.
\par
According to a theorem by Lovelock \cite{lovelock71},
GR is the only local, generally covariant theory in four-dimensional spacetime, derivable from an action, with a single metric as the gravitational degree of freedom, whose field equations are of second order, and that has special relativity as its no-gravity  limit.
A relativistic extension of MOND, as any other generalization of GR, must then break with one of these restrictions.
\par
In MOND, which introduces a fiducial constant with the dimensions of acceleration, this fact is further brought home to us for the following reason: In constructing a relativistic action, we need to identify a quantity constructed from the degrees of freedom -- a scalar, if we want the theory to be covariant -- that has the dimensions of acceleration. The theory then compares the value of this scalar with $\az$ for the system under study and directs us according to whether we are in the high- or low-acceleration regime.\footnote{In analogy with quantum theory, which introduces $\hbar$ of the dimensions of action, where system attributes of these dimensions are to be compared with $\hbar$ to indicate whether we are in the quantum or classical regime.} In modified gravity theories, such as the one I deal with here, where we modify the gravitational action, we need to construct this quantity from the gravitational degrees of freedom. But, we cannot construct such a scalar from a single metric and its first derivatives.
\par
We can, for example, relinquish locality, and write acceleration scalars from nonlocal differential operations on the metric, as in Refs. \cite{soussa03,deffayet14}. Or we can relinquish general covariance, invoking a preferred frame, as in Ref. \cite{milgrom19a}, which, however, is equivalent to a covariant theory with added degrees of freedom -- a so-called  $f(Q)$ theory \cite{milgrom19a,dambrosio20}.
\par
Or else, we can break off the Lovelock requirements by introducing additional gravitational degrees of freedom. For example, the first (toy) relativistic MOND theory \cite{bm84} is a scalar-tensor theory, discarded because it does not produce single-potential lensing in the above sense. The first relativistic MOND theory with single-potential lensing was put forth by Sanders in Ref. \cite{sanders97}. But it included a nondynamical, Lorentz-symmetry-breaking background vector field. Bekenstein \cite{bekenstein04} then turned this into a full fledged MOND theory with single-potential lensing called TeVeS. TeVeS with its various versions (characterized by different couplings between the tensor, vector, and scalar degrees of freedom) has been the most widely studied relativistic MOND theory.
\par
Particularly noteworthy is the recent advent of a TeVeS version dubbed RMOND  (for relativistic MOND) \cite{sz19,sz21,sz21a} (see also  Ref. \cite {kashfi22}). This theory, beyond reproducing NR MOND phenomenology (including lensing), is said to imply that the propagation speed of tensor gravitational waves equals that of light under all circumstances, and it can also reproduce the observed properties of the microwave background radiation and of the formation of large-scale structure.
\par
Bimetric MOND (BIMOND) \cite{milgrom09b} -- which is the subject of this paper -- is a class of relativistic MOND formulations in which the gravitational degrees of freedom are two metrics: one, $\gmn$, that couples to ``matter,'' which we perceive as our world, and another metric, $\hgmn$, that couples to ``twin matter,'' and which is assumed not to interact directly with matter.
\par
The availability of a second metric enables one to construct ``relative-acceleration'' scalars: The affine connections of the two metrics\footnote{Taken here as the metric-compatible, symmetric, Levi-Civita-Christoffel connections --  $\C\ten{\l}{\m\n}$ and $\hat\C\ten{\l}{\m\n}$ respectively} are not themselves tensors, but their difference,
\beq C\ten{\l}{\m\n}=\C\ten{\l}{\m\n}-\hat\C\ten{\l}{\m\n}, \eeqno{jamaja}
is a tensor.\footnote{The tensor $C\ten{\l}{\m\n}$ may be thought of as the ``relative acceleration field'' of $\gmn$ with respect to $\hgmn$, in the sense that in a frame where $\hgmn$ is locally flat at some $\hat x$ [i.e., $\hat\C\ten{\m}{\a\b}(\hat x)=0$],
 we have $d\^2x\^\m/d\tau \^2|\_{\hat x}=-C\ten{\m}{\a\b}\dot x\^\a\dot x\^\b$, on a geodesic of $\gmn$ through $\hat x$.}
This relative-acceleration tensor can be used to construct various ``acceleration scalars'' by various contractions of powers of $C\ten{\l}{\m\n}$.
From these, one can construct terms in a BIMOND action that represent interactions between the two metrics, and one can choose these interactions so as to give MOND phenomenology in the NR limit.
\par
We also want to retain all the successes of GR in the high-acceleration regime. To achieve this, BIMOND is constructed by adding the BIMOND interaction term to the standard Einstein-Hilbert actions of the two metrics, and decreeing that this interaction vanishes in the high-acceleration limit.
\par
The general, schematic form of the BIMOND gravitational action is thus
\beq I=-\frac{1}{\spg}\int d^4x ~[A|g|\^{1/2}R+\hat A|\hat g|\^{1/2}\hat R+v(g,\hat g)\lM\^{-2}\M(\{\lM C\})],  \eeqno{moira}
where $v(g, \hat g)$ stands for a volume element constructed from the two metrics, $\{\lM C\}$ in the argument of the interaction Lagrangian, $\M$, stands for a collection of scalars constructed from the dimensionless $\lM C\ten{\l}{\m\n}$, and $\lM=\az^{-1}$ is the MOND length, where I use units where $c=1$, adopted hereafter.
\par
For some reasons -- unjustified as I now realize -- all previous studies of BIMOND's various aspects \cite{milgrom09b,milgrom10b,milgrom10a,milgrom14b,clifton10,milgrom19a,dambrosio20} dealt with a BIMOND version that employs essentially a single scalar, obtained by contracting the tensor
\beq  \Up\_{\m\n}\equiv C\ten{\c}{\m\l}C\ten{\l}{\n\c}-C\ten{\l}{\m\n}C\ten{\a}{\l\a}  \eeqno{upup}
with $\Gmn$ or $\hGmn$, namely, $\Up\equiv\Gmn\Up\_{\m\n}$ and/or $\hat\Up\equiv\hGmn\Up\_{\m\n}$. These two scalars define in a sense only one variable of $\M$ because they represent the same basic scalar (see Sec. \ref{choice}). So in this formulation, the effective BIMOND Lagrangian is a function of a single variable $\Z\_{\Up}=\lM\^2\Up$.
The form of $\M(\Z)$ is then determined from MOND phenomenology in the NR limit (it is tantamount to the so-called MOND interpolating function). Thus in such limited BIMOND versions, what we know from the NR phenomenology fully determines the theory also in the relativistic regime.
\par
The main result of this work is that there is a much larger subclass of BIMOND theories that have a single-potential NR MOND phenomenology. Instead of one, there are several independent quadratic scalars (and all their linear combinations) that yield such a phenomenology. The interaction Lagrangian may thus be taken as a multivariable function of these, $\M(\Z_1,\Z_2,...)$. In the NR limit, however, all these scalars tend to the same
NR scalar acceleration variable, $\bar\Z$, and the interaction reduces to a function of this single variable, $\bar\M(\bar\Z)$, yielding the NR, MOND interpolating function, which is determined by NR phenomenology. As a consequence, knowledge of $\bar\M(\bar\Z)$ constrains the full  $\M(\Z_1,\Z_2,...)$ only very poorly. For example, there are acceleration scalars, $\Z$, that do not contribute at all in the NR limit, but may be important beyond it, and so, the dependence of the theory on them disappears altogether in the NR limit.
\par
Here, I ignore the important questions of stability properties and the possible existence of ghosts in such theories. These theories are in any event, at best, effective theories that clearly do not encapsule the whole story. So perhaps even theories that on the face of it are unstable may be cured by adding elements. In particular, this may be so in the context of MOND, where departures from (the stable) GR are characterized by a MOND length $\lM\sim \ell\_U$ and a MOND time of the order of the Hubble time.
\par
As shown in Ref. \cite{milgrom19a}, BIMOND, with the ``auxiliary'' metric constrained to be flat, is equivalent to so-called $f(Q)$ formulations of MOND. So the present study also enlarges the class of MOND $f(Q)$ theories with a good NR limit.
\par
Even if BIMOND itself turns out to be lacking in some regards (for example, if it is not able to describe the observed cosmology in full), it is useful to study it as a possible guide to better theories.
\par
I note, finally, that all full-fledged, relativistic MOND theories propounded to date (including the one I discuss here), might be described as ``modified gravity,'' in the sense that they involve a modification of the Einstein-Hilbert action -- the ``gravitational,'' or the ``free'' action of gravity in GR -- without modifying the matter actions. The latter comprise the free (inertial) matter actions, and the interactions between matter degrees of freedom. There might be other routes to a fundamental MOND theory, involving, for example, modifications of the free matter actions (perhaps all parts of the action). These go under the name of ``modified inertia.'' Only preliminary suggestions in this vein have been considered so far, and only for the NR regime (see the recent Ref. \cite{milgrom22}).
\par
In Sec. \ref{choice}, I discuss in some detail the scalars that we can form from powers of $C\ten{\l}{\m\n}$.
In Sec. \ref{field}, I give the detailed action and derive the general field equations.
In Sec. \ref{nrscalars}, I derive the NR limit and identify the general quadratic BIMOND scalars that ensure a single-potential NR limit.
In Sec. \ref{flrw}, I discuss a class of cosmological solutions whereby the two metrics are small departures from a single Friedman-Lemaitre-Robertson-Walker (FLRW) geometry.
\section{The BIMOND interaction scalars  \label{choice}}
We are interested in constructing scalars by contracting products of $C\ten{\a}{\b\c}$ (or polynomials of them) with $\gmn,~\Gmn,~\hgmn,~\hGmn$ and $\d\ten{\m}{\n}$, and with coefficients that are functions of the scalars $\k=(g/\hat g)\^{1/4}$ and $\bo=\Gmn\hgmn$.
\par
Since the $C\ten{\l}{\m\n}$ have three indices, and the contracting tensors at our disposal have an even number of indices, we can construct only scalars from an even power of the $C\ten{\l}{\m\n}$.\footnote{Higher powers can also be contracted with $|g|\^{1/2}\eps\_{\a\b\c\z}$, etc.}
\par
But, for the treatment to be more manageable, we shall restrict ourselves to scalars that are quadratic in the $C\ten{\l}{\m\n}$. Since I will consider functions of such scalars, this would include, in effect, dependence on higher-order scalars that are products of quadratic ones. But this is still a substantial restriction, because, clearly, one can construct higher-order scalars that are not powers of quadratic ones, for example
$\Gmn\hGab C\ten{\l\_1}{\a\l\_4}C\ten{\l\_2}{\b\l\_1}C\ten{\l\_3}{\m\l\_2}C\ten{\l\_4}{\m\l\_3}$.
\par
The general quadratic scalars are of the form
 \beq S=Q\ten{\b\c\m\n}{\a\l}C\ten{\a}{\b\c}C\ten{\l}{\m\n},
   \eeqno{cures}
where $Q\ten{\b\c\m\n}{\a\l}$ is a linear combination of terms built from $\gmn$, $\hat g\_{\m\n}$; their inverses, $\d\ten{\a}{\b}$; and the scalars $\k$ and $\bo$.
We can write this as
\beq S=\Qt(\gt,\hgt)\Ct\Ct,  \eeqno{hytsc}
where boldface denotes a tensor. In expression (\ref{hytsc}) all indices are understood to be contracted.
\par
There are only five independent, basic forms of  $\Qt(\gt,\hgt)$. Their components can be chosen as
$$ {\Qt_1(\gt,\hgt)}\ten{\b\c\m\n}{\a\l}=[\grp\^{\b\m}\drp\ten{\n}{\a}\drp\ten{\c}{\l}]_s,~~~ {\Qt_2(\gt,\hgt)}\ten{\b\c\m\n}{\a\l}=[\grp\^{\b\c}\drp\ten{\m}{\a}\drp\ten{\n}{\l}]_s,~~~
{\Qt_3(\gt,\hgt)}\ten{\b\c\m\n}{\a\l}=[\grp\_{\a\l}\grp\^{\b\c}\grp\^{\m\n}]_s,$$
\beq {\Qt_4(\gt,\hgt)}\ten{\b\c\m\n}{\a\l}=[\grp\^{\b\m}\drp\ten{\c}{\a}\drp\ten{\n}{\l}]_s,~~~
 {\Qt_5(\gt,\hgt)}\ten{\b\c\m\n}{\a\l}=[\grp\_{\a\l}\grp\^{\b\m}\grp\^{\c\n}]_s,  \eeqno{meutaz}
where $\grp$ stands for either $g$ or $\hat g$, and $\drp\ten{\a}{\b}$ stands for one of the following: $\d\ten{\a}{\b}$, $q\ten{\a}{\b}\equiv g\^{\a\s}\hat g\_{\s\b}$,  or its inverse $\hat g\^{\a\s} g\_{\s\b}$.
Also, $[~~]_s$ signifies that the expression is symmetrized under $\m\leftrightarrow\n$, under $\b\leftrightarrow\c$, and under $(\a,\b,\c)\leftrightarrow (\l,\m,\n)$.
The general $\Qt$ is a linear combination of such tensors, with coefficients that can depend on the scalars $\k$ and $\bo$.
\par
Note that in a given expression any of the symbols can stand for any of their meanings; so, for example, the two appearances of $\drp\ten{\a}{\b}$
in the expression for $\Qt_1$ can have different meanings.
\par
As will be discussed below, there are situations in which the two metrics differ by only a small amount, such that to a good enough approximation, we can replace both metrics by a single reference metric, keeping the difference only in the relative acceleration tensor $C\ten{\l}{\m\n}$
and the scalar derived from it. In this case all the contractions give the same quantity. This applies, e.g., to the NR limit, and more generally, to the weak-field limit on a background of double Minkowski.
\par
Instead of proceeding with the more cumbersome general case of arbitrary contraction schemes for the scalar, for clarity of exposition I shall give subsequent expressions for the case where
only one of the metrics at a time is used for contractions, namely, I use $\Qt(\gt,\gt)$ to form the basic scalars. This is straightforward to generalize. The basic scalars can then be taken as
\beq S_1\equiv\Gmn C\ten{\c}{\m\l}C\ten{\l}{\n\c},~~~S_2\equiv\bar C\^{\c}C\_{\c},~~~
 S_3\equiv\gmn\bar C\^{\m}\bar C\^{\n},~~~ S_4\equiv\Gmn C\_{\m} C\_{\n},
 ~~~S_5\equiv g\_{\a\l}g\^{\b\m} g\^{\c\n}C\ten{\a}{\b\c}C\ten{\l}{\m\n},
 \eeqno{meqaz}
  where
$\bar C\^{\c}\equiv \Gmn C\ten{\c}{\m\n}$ and $C\_{\c}\equiv C\ten{\a}{\c\a}$  are the two traces of $C\ten{\l}{\m\n}$.\footnote{Derivatives of the scalars $\k$ and $\bo$ are expressible in terms of $C\ten{\a}{\b\c}$:  $\k\_{,\n}=(1/2)\k C\_{\n}$, and $\bo\der{\n}=-2q\ten{\m}{\a}C\ten{\a}{\m\n}$.}

\subsection{Symmetrization with respect to metric interchange\label{interchange}}
There are interesting BIMOND theories that do not treat the two metrics in a symmetric way. For example, the MOND theory described in Ref. \cite{milgrom19a}, as well as $f(Q)$ formulations of MOND \cite{milgrom19a,dambrosio20} are equivalent to a metric-asymmetric BIMOND, with the auxiliary metric constrained to be flat.
But, symmetric theories have their appeal, and restricting ourselves to treating them, as I do henceforth, reduces the treatment to more manageable proportions.
\par
Such a symmetry of the theory can be achieved in different ways. For example, we can employ scalars that are themselves not necessarily symmetric to metric interchange, but symmetrize the theory by taking with each term in the Lagrangian a term that is gotten from it by metric interchange.
Alternatively, we can construct the Lagrangian from scalars that have built-in metric-exchange symmetry, which is how I proceed here, for the sake of concreteness.
\par
Since $C\ten{\a}{\b\c}$ is antisymmetric to metric interchange, a product of an even number of them is interchange symmetric. But to form scalars we have to contract the indices using the two metrics and their inverses. A product of two $C$s cannot be contracted with the metrics  in a symmetric way because it presents two upper and four lower indices to contract.
\par
Thus, I have in mind linear combinations of the symmetrized
\beq \oot[\Qt(\gt,\hgt)+\Qt(\hgt,\gt)], \eeqno{rabasar}
for example, we can take as symmetrize basic scalars
\beq \Qt_i(\gt,\hgt)=\oot[\Qt_i(\gt,\gt)+\Qt_i(\hgt,\hgt)]~~~~(i=1,...,5) \eeqno{rabuter}
with $\Qt_i(\gt,\gt)$ corresponding to the scalars in Eq. (\ref{meqaz}).
\par
In what follows I shall speak of $S_i$ with the understanding that they are symmetrized with respect to metric interchange.
\par
Having said that, in the application of BIMOND I shall consider in this paper, the two metrics will be small departures from a common reference metric $\rgt$, and I shall consider the lowest order in these departures. $C\ten{\a}{\b\c}$ vanish when the metrics are equal and are of first order in their small difference. Thus, to lowest order only $\Qt(\rgt,\rgt)$ will be used to construct the scalars, and this gives, automatically, interchange-symmetric scalars. So the whole issue of the symmetry will not really be brought to bear here.

\section{The theory \label{field}}
After all the restrictions described above -- made essentially for the sake of concreteness and tractability -- I consider the following BIMOND action, symmetric in the two metrics, and employing only quadratic scalars:
\beq I=I\_G(\gmn,\hgmn)+I\_M(\gmn,\Psi)+\hat I\_M(\hgmn,\hat\Psi),  \eeqno{action}
where $I\_M$ and $\hat I\_M$ are, respectively the matter and ``twin matter'' actions, in which the respective metrics are assumed to couple minimally to the matter degrees of freedom, schematically denoted $\Psi$ and $\hat\Psi$.
The gravitational action is taken as
\beq I\_G=-\frac{1}{\spg}\int d^4x ~\{|g|\^{1/2}R+|\hat g|\^{1/2}\hat R+(g\hat g)\^{1/4}\az^2[\M(\Z_1,\Z_2,...)+\Z\_{\Up}]\},  \eeqno{muba}
where the first two terms are the standard Einstein-Hilbert actions for the two sectors,\footnote{It is possible to couple the two metrics also in the Einstein-Hilbert terms, not through the $\Z_m$ scalars (such as through the volume elements) but I avoid such couplings, since I want the theory to go to GR in the limit $\az\rar\infty$.}
the third term is the interaction action,
and $\Z_m=S_m/\azs$, where $S_m$ are quadratic scalars.\footnote{In our units ($c=1$), $\az=\lM\^{-1}$ has dimensions of inverse length.} Hence, $\Z_m$ are of the form
\beq\Z=\az\^{-2}Q\ten{\b\c\m\n}{\a\l}C\ten{\a}{\b\c}C\ten{\l}{\m\n}.\eeqno{zababa}
The scalar $\Z\_{\Up}\equiv(1/2)(\Gmn+\hGmn)\Up\_{\m\n}/\azs$ -- also of the form (\ref{zababa}) -- is the symmetrized $\Up_s/\azs$ scalar used in the earlier BIMOND studies; it is added to the Lagrangian density for later convenience. For concreteness' sake, I work with the volume element $v(g,\hat g)=(g\hat g)\^{1/4}$.
\par
In constructing a MOND theory we want to retain the successes of GR in the high-acceleration regimes, as was done in constructing GR, given the successes of Newtonian dynamics, and in constructing quantum mechanics, given the successes of classical mechanics. Thus, a ``correspondence principle''  is adopted, whereby the Lagrangian density in square brackets goes to a constant (of order unity) in the limit $\az\rar 0$. Each sector is then equivalent to GR with a cosmological constant $\sim\azs$. With our specific choice of volume element, the two sectors remain somewhat coupled through the ``cosmological-constant'' term. For example, the value of the cosmological constant  measured in the $\gmn$ sector is $\sim (\hat g/g)\^{1/4}\azs$, and that in the twin sector is $\sim (g/\hat g)\^{1/4}\azs$. The decoupling in the high-acceleration limit can be made complete by taking, e.g., a volume element such as $v(g,\hat g)=(|g|\^{1/2}+|\hat g|\^{1/2})/2$.
\par
Varying over $\gmn$ and $\hgmn$ gives, respectively, the field equations
\beq G\^{\m\n}+\T\^{\m\n}+\epg T\^{\m\n}\_{M}=0,   \eeqno{melte}
\beq \hat G\^{\m\n}+\hat \T\^{\m\n}+\epg \hat T\^{\m\n}\_{M}=0,   \eeqno{melteha}
where
$G\^{\m\n}$ and $\hat G\^{\m\n}$ are the Einstein tensors of the two metrics, $T\^{\m\n}\_M$ and $\hat T\^{\m\n}\_M$ are the matter and twin-matter energy-momentum tensors (EMT), and $\T\^{\m\n}$ and $\hat \T\^{\m\n}$ are those gotten from varying the BIMOND interaction term. Their normalization is such that varying the interaction action with respect to $\gmn$ is
\beq \d I\_I=\frac{1}{\spg}\int d\^4x~(-g)\^{1/2}\T\^{\m\n}\d\gmn, \eeqno{bubure}
and similarly for the other sector.
\par
According to our ``correspondence principle,'' in the high acceleration limit $\T\^{\m\n}$ and $\hat \T\^{\m\n}$ are dark-energy-like contributions of order $\azs$.
\par
The Bianchi identities, and the related Cauchy problem for BIMOND are discussed in Ref. \cite{milgrom09b}: While $G\^{\m\n}$ and $\hat G\^{\m\n}$ are identically divergenceless (each with its own covariant divergence), $\T\^{\m\n}$ and $\hat \T\^{\m\n}$ are not identically divergenceless. But they are so for solutions of the field equations, since  the matter actions are. There are, however, four combined Bianchi identities following from the coordinate covariance of the interaction action. They are
\beq |g|\^{1/2}{\T\^{\m\n}}\_{;\n}+|\hat g|\^{1/2}{{\hat\T}\^{\m\n}}\_{~~~:\n}=0.   \eeqno{bianch}
Here, ``;''  denotes the covariant derivative with respect to $\gmn$, and ``:''  that with respect to $\hgmn$.

\subsection{Derivation of $\T\^{\m\n}$ and $\hat \T\^{\m\n}$}

We identify three contributions to $\T\^{\m\n}$, denoted and discussed separately, because they turn out to play different roles in the dynamics.
\par
The variation over $\gmn$ in the volume element contributes to $\Tmn$
\beq \T\ten{(1)}{\m\n}=-\frac{1}{4} (\hat g/g)\^{1/4}[\az^2\M(\Z_m)+\Up_s]\gmn, \eeqno{con1}
and symmetrically for $\hgmn$,
 \beq \hat \T\ten{(1)}{\m\n}=-\frac{1}{4} (g/\hat g)\^{1/4}[\az^2\M(\Z_m)+\Up_s]\hgmn. \eeqno{con1hat}
 (It is more convenient to use lower indices on the EMT; indices are lowered in each sector with the corresponding metric: $\Tmn=g\_{\m\a}g\_{\n\b}\T\^{\a\b}$, $\hTmn=\hat g\_{\m\a}\hat g\_{\n\b}\hat\T\^{\a\b}$.)
\par
Varying over $\gmn$ as it appears in the $Q\ten{\b\c\m\n}{\a\l}$, gives a contribution  to $\Tmn$ of the form
\beq \T\ten{(2)}{\m\n}=(\hat g/g)\^{1/4}[-\sum_m\pd{\M}{\Z_m}\U\_{\m\n}\^{(m)}+\oot\Up\_{\m\n}],  \eeqno{con2}
where $\U\_{\m\n}(m)$ are expressions quadratic in $C\ten{\a}{\b\c}$, with two free indices.
Because we want to separate factors that are explicitly proportional to metric differences,
write it schematically as $\U\_{\m\n}=[u(\gt,\hgt)C^2]\_{\m\n}$.

\par
Similarly, when varying with respect to $\hgmn$, we get the contribution
\beq \hat \T\ten{(2)}{\m\n}=(g/\hat g)\^{1/4}[-\sum_m\pd{\M}{\Z_m}\hat\U\_{\m\n}\^{(m)}+\oot\Up\_{\m\n}],  \eeqno{con2hat}
where, from the assumed symmetry to metric interchange, $\hat \U\_{\m\n}(\gt,\hgt)=[u(\hgt,\gt)C^2]\_{\m\n}$.
\par
When a single metric, say $\gmn$, is used to construct $Q\ten{\b\c\m\n}{\a\l}$, we have for the basic scalars themselves
$$ \U\ten{(1)}{\b\c}=- C\ten{\s}{\c\k}C\ten{\k}{\b\s},~~~~\U\ten{(2)}{\b\c}=-C\ten{\s}{\b\c}C\ten{\a}{\s\a}=-C\ten{\s}{\b\c}C\_{\s},~~~~\U\ten{(3)}{\b\c}=g\_{\b\l}\Gmn g\_{\c\r}g\^{\a\s}C\ten{\l}{\m\n} C\ten{\r}{\a\s}-2g\_{\a\l}\Gmn C\ten{\l}{\m\n}C\ten{\a}{\b\c}=g\_{\b\l} g\_{\c\r}\bar C\^\l \bar C\^\r-2g\_{\a\l}\bar C\^\l C\ten{\a}{\b\c},$$
\beq\U\ten{(4)}{\b\c}=-C\ten{\a}{\b\a} C\ten{\l}{\c\l}=-C\_\b C\_\c,~~~~ \U\ten{(5)}{\b\c}=g\^{\a\l}g\^{\s\r}g\_{\b\eta}g\_{\c\d}C\ten{\eta}{\a\s}C\ten{\d}{\l\r}-2g\_{\a\l}g\^{\s\r}C\ten{\a}{\b\s}C\ten{\l}{\c\r}. \eeqno{mugaga}
If we symmetrize the scalar by employing $\Qt(\hgt,\gt) = [\Qt(\gt,\gt)+\Qt(\hgt,\hgt)]/2$, the corresponding
$\U\_{\b\c}(\hgt,\gt) = [\U\_{\b\c}(\gt,\gt)+\U\_{\b\c}(\hgt,\hgt)]/2$ is a similar symmetrized form of expressions (\ref{mugaga}). ($\U\ten{(1)}{\b\c}$, $\U\ten{(2)}{\b\c}$, and $\U\ten{(4)}{\b\c}$ involve no contractions with a metric, so they are the same in the two sectors, and so is $\Up\_{\m\n}$.)
\par
The third contribution comes from varying over the metrics as they appear in the $C\ten{\a}{\b\c}$, which involve both the metrics and their derivatives.
Starting from the $\M$ term in the Lagrangian:
 Varying over $\gmn$, and concentrating on one of the variables for clarity,
\beq \d I\^{(3)}=-(\spg)\^{-1}\int d\^4x~(g\hat g)\^{1/4}\M'(\Z)Q\ten{\b\c\m\n}{\a\l}(C\ten{\a}{\b\c}\d C\ten{\l}{\m\n}+ C\ten{\l}{\m\n}\d C\ten{\a}{\b\c})=-(\epg )\^{-1}\int d\^4x~(g\hat g)\^{1/4}\M'(\Z)Q\ten{\b\c\m\n}{\a\l}C\ten{\l}{\m\n}\d \C\ten{\a}{\b\c},   \eeqno{karupas}
where I used the symmetry of $Q\ten{\b\c\m\n}{\a\l}$ to the interchange $(\a,\b,\c)\leftrightarrow (\l,\m,\n)$.
Now,
\beq \d \C\ten{\a}{\b\c}=\oot\d g\^{\a\s}(g\_{\b\s,\c}+g\_{\c\s,\b}-g\_{\b\c,\s})+\oot g\^{\a\s}(\d g\_{\b\s,\c}+\d g\_{\c\s,\b}-\d g\_{\b\c,\s}). \eeqno{nunuma}
$\d \C\ten{\a}{\b\c}$ is a tensor which equals $\oot g\^{\a\s}(\d g\_{\b\s;\c}+\d g\_{\c\s;\b}-\d g\_{\b\c;\s})$ in a locally flat system (of $\gmn$, with respect to which the covariant derivative is taken); so they are equal in any frame.
\par
Take the first term, and write the integral as
\beq \int  d\^4x~(-g)\^{1/2}A\ten{\b\c}{\a}g\^{\a\s}\d g\_{\b\s;\c}=\int  d\^4x~(-g)\^{1/2}(A\ten{\b\c}{\a}g\^{\a\s}\d g\_{\b\s})\_{;\c}-\int d\^4x~ (-g)\^{1/2}(A\ten{\b\c}{\a}g\^{\a\s})\_{;\c}\d g\_{\b\s},  \eeqno{lakalaka}
where $A\ten{\b\c}{\a}= (\hat g/g)\^{1/4}\M'(\Z)Q\ten{\b\c\m\n}{\a\l}C\ten{\l}{\m\n}$.
The first integrand is a divergence of a vector  [$(-g)\^{1/2}(A\ten{\b\c}{\a}g\^{\a\s}\d g\_{\b\s})\_{;\c}=(A\ten{\b\c}{\a}g\^{\a\s}\d g\_{\b\s})\der{\c}$] and gives a surface integral that vanishes, and similarly for the other terms.
\par
The contribution from the $\Up_s$ part of the Lagrangian:
\beq \T\ten{(3,\Up)}{\m\n}=-\oot\{(\hat g/g)\^{1/4}[C\ten{\l}{\l(\m}\d\ten{\c}{\n)}-C\ten{\c}{\m\n}-\oot\gmn(g\^{\c\a}C\ten{\l}{\a\l}-\Gab C\ten{\c}{\a\b})]\}\cd{\c}, \eeqno{matayata}
where $(\m\n)$ indicates symmetrization over $\m\n$: $A\_{(\m...\n)}\equiv (1/2)(A\_{\m...\n}+A\_{\n...\m})$.
\par
Putting all together, and returning the sum over the variables, we have
\beq \T\ten{(3)}{\m\n}=\{(\hat g/g)\^{1/4}[\sum_m\pd{\M}{\Z_m}S\^{\c}\_{\m\n}(m)-\oot C\ten{\l}{\l(\m}\d\ten{\c}{\n)}+\oot C\ten{\c}{\m\n}+\frac{1}{4}\gmn(g\^{\c\a}C\ten{\l}{\l\a}-\Gab C\ten{\c}{\a\b})]\}\_{;\c},  \eeqno{con3}
where
\beq S\ten{\c}{\m\n}=(g\_{\a\m}[Q\ten{\c\a\b\s}{\n\l}]_s+g\_{\a\n}[Q\ten{\c\a\b\s}{\m\l}]_s-g\^{\a\c}g\_{\r\m}g\_{\z\n}[Q\ten{\r\z\b\s}{\a\l}]_s)C\ten{\l}{\b\s}.   \eeqno{nanatata}
The notation $[~~]_s$ signifies that the $Q$'s have to be symmetrized over the first two upper indices, the second two upper indices, and the interchange of the first three with the last three (two upper and one lower).
\par
Similarly, for variation over $\hgmn$ we get, by interchanging the metrics,
\beq \hat \T\ten{(3)}{\m\n}=-\{(g/\hat g)\^{1/4}[\sum_m\pd{\M}{\Z_m}\hat S\^{\c}\_{\m\n}(m)-\oot C\ten{\l}{\l(\m}\d\ten{\c}{\n)}+\oot C\ten{\c}{\m\n}+\frac{1}{4}\hgmn(\hat g\^{\c\a}C\ten{\l}{\a\l}-\hGab C\ten{\c}{\a\b})]\}\_{:\c},  \eeqno{con3hat}
where in $\hat S\^{\c}\_{\m\n}$, $\hgmn$ replaces $\gmn$.
The minus sign in the expression for $\hat \T\ten{(3)}{\m\n}$ relative to that for $\T\ten{(3)}{\m\n}$ is due to the fact that $C\ten{\a}{\b\c}$ changes sign under metric interchange, while $Q$ is symmetric.
\par
$S\^{\c}\_{\m\n}(m)$ and $\hat S\^{\c}\_{\m\n}(m)$ correspond to the $m$th scalar variable $\Z_m$, and are linear combinations of those for the individual basic scalars, for which we have, for the case where a single metric, say $\gmn$, is used to construct $Q\ten{\b\c\m\n}{\a\l}$,
$$ S\ten{(1)\c}{\m\n}=C\ten{\c}{\m\n},~~~~~S\ten{(2)\c}{\m\n}=\oot\gmn(\Gab C\ten{\c}{\a\b}-g\^{\c\a}C\ten{\l}{\a\l})+\d\ten{\c}{(\m}C\ten{\l}{\n)\l},~~~~~S\ten{(3)\c}{\m\n}=-\gmn\Gab C\ten{\c}{\a\b}+2\d\ten{\c}{(\m}g\_{\n)\r}\Gab C\ten{\r}{\a\b},$$
\beq S\ten{(4)\c}{\m\n}=\gmn g\^{\c\a}C\ten{\l}{\a\l},~~~~~S\ten{(5)\c}{\m\n}=-C\ten{\c}{\m\n}+2g\^{\c\s}g\_{\l(\m}C\ten{\l}{\n)\s}.\eeqno{malimali}
If both metrics are used in the contraction of a given scalar, then these expressions need to be somewhat modified.
For example, if symmetrization of a scalar is applied by defining $Q(g,\hat g)=\oot[Q(g,g)+Q(\hat g,\hat g)]$, then expressions (\ref{malimali}) need to be similarly symmetrized only over the explicit appearance of the metric (not in $C\ten{\a}{\b\c}$).
\par
In the high-acceleration limit, $\az\rar 0$, we have $\T\^{\m\n}(2),~\hat \T\^{\m\n}(2),~\T\^{\m\n}(3),~\hat \T\^{\m\n}(3)\rar 0$, and $\T\^{\m\n}(1),~\hat \T\^{\m\n}(1)$ is dark energy of order $\azs$, by our correspondence principle.

\section{The nonrelativistic limit \label{nrscalars}}
The weak-field limit (WFL) of BIMOND, to be discussed in more detail in Sec. \ref{wflscalars}, concerns systems where the metrics depart only a little from a common Minkowski metric, in which limit, the departure from Minkowski is treated to lowest order. In the important special case of the nonrelativistic (NR) limit, the gravitational field is also static, or time independent; to wit, the field varies on timescales much longer than the light crossing time over the system under study. This means, in particular, that the sources of gravity in the system are slowly moving -- static in the limit. Their energy-momentum tensor has only a time-time component equal to the mass density, and the metric is taken to be time independent.
\par
This situation applies to the dynamics in galactic systems, including the description of gravitational lensing by galactic systems.
\par
Following Ref. \cite{milgrom09b}, write for the weak-field limit
 \beq g\_{\m\n}=\emn-2\f\d\_{\m\n} +h\_{\m\n},~~~~
 \hat g\_{\m\n}=\emn-2\fh\d\_{\m\n}+\hat h\_{\m\n}.
\eeqno{rutza}
Defining $\f\equiv (\eta\_{00}-g\_{00})/2$ and  $\fh\equiv (\eta\_{00}-\hat g\_{00})/2$, we have $h\_{00}=\hat h\_{00}=0$.
Also, denote $e\_i\equiv h\_{0i}=\hat h\_{0i}$ (Roman letters are used for space indices).
\par
 We denote the differences and sums of the potentials
 \beq \fs=\f-\fh, ~~~\hs\_{\m\n}=h\_{\m\n}-\hat h\_{\m\n}~~~\f\^+=\f+\fh, ~~~h\^+\_{\m\n}=h\_{\m\n}+\hat h\_{\m\n}.
 \eeqno{metza}

\par
We wish to solve the field equations to first order in the potentials
$\f,~\fh,~h\_{ij},~ \hat h\_{ij},~e\_i,~\hat e\_i$ (I call them $\psi$ when I refer to them collectively).
\par
The dominant, lowest-order terms in the field equations are of the order $\partial^2\psi$. A close inspection shows that the $\T\ten{(1)}{\m\n}$ and $\T\ten{(2)}{\m\n}$ terms can be neglected in our approximation. They are either of order $(\partial\psi)^2$, which are of order $\psi\ll 1$ relative to the dominant term, or they are of order $\azs$, which contribute a density of the order of the dark-energy density [by Eq. (\ref{coinc})], which we neglect in the context of local systems.
\par
As to the $\T\ten{(3)}{\m\n}$ terms, we can replace in them the metrics themselves -- but not their derivatives -- by $\emn$. We also replace the covariant derivatives by derivatives, because the correction to this is also of order $(\partial\psi)^2$.
This leaves us with
\beq -\hat\T\ten{(3)}{\m\n}\approx\T\ten{(3)}{\m\n}\approx\bar\T\ten{(3)}{\m\n}\equiv[\sum_m\pd{\M}{\Z_m}S\^{\c}\_{\m\n}(m)-\oot C\ten{\l}{\l(\m}\d\ten{\c}{\n)}+\oot C\ten{\c}{\m\n}+\frac{1}{4}\emn(\eta\^{\c\a}C\ten{\l}{\l\a}-\eta\^{\a\b} C\ten{\c}{\a\b})]\der{\c},  \eeqno{con3nr}
where in $\Z_m$ and $S\^{\c}\_{\m\n}(m)$ all contractions and raising and lowering indices are done with $\emn$.
\par
It is useful to consider the sum and the difference of the field equations for the two sectors.
In the WFL we then have
\beq G_{\m\n}+\hat G_{\m\n}=-\epg(T\^M\_{\m\n}+\hat T\^M\_{\m\n}),  \eeqno{sumasa}
and
\beq G_{\m\n}-\hat G_{\m\n}+2\bar\T\ten{(3)}{\m\n}=-\epg(T\^M\_{\m\n}-\hat T\^M\_{\m\n}),  \eeqno{difasa}
\par
The WFL, Einstein tensors are linear in $\partial\psi$; so the sum equation of the WFL of BIMOND is equivalent to the WFL of GR for the potentials $\psi\^+\equiv\psi+\hat\psi$. This sum sector also enjoys the full gauge freedom; thus its solutions are those of GR.
\par
As to the difference sector, $G_{\m\n}-\hat G_{\m\n}$ is linear in $\partial^2\psi\^-$ (where $\psi\^-=\psi-\hat\psi$); specifically, to lowest order
\beq  \d G\_{\m\n}=G_{\m\n}-\hat G_{\m\n}\approx\d R\_{\m\n}-\oot\emn\eta\^{\a\b}\d R\_{\a\b}=[C\_{(\m}\d\ten{\c}{\n)}-C\ten{\c}{\m\n}-\oot\emn(C\^\c-\bar C\^\c)]\der{\c}. \eeqno{oioioi}
($C\ten{\l}{\m\n}$ are linear in $\partial\psi\^-$.)
\par
Substituting this and expression (\ref{con3nr}) in Eq. (\ref{difasa}) gives the difference equation
\beq [\sum_m\pd{\M}{\Z_m}S\ten{\c}{\m\n}(m)]\der{\c}=-\fpg(T\^M\_{\m\n}-\hat T\^M\_{\m\n}).\eeqno{surtino}
All the above is valid in the general WFL.
\par
Specialize now to the NR limit, where all the potentials are time independent, and the NR matter energy-momentum tensors are $T\^M\_{\m\n}=\r\d\_{\m 0}\d\_{\n 0}$, and similarly for the hatted one.
The sum sector is equivalent to GR for the sum of potentials, and is sourced by $\r+\rh$. Also, as in GR, there is a gauge where $\hmn\^+=0$, and $\f+\fh$ is the solution of the Poisson equation sourced by $\r+\rh$.
\par
For the difference field equation, we have to the desired order,
 $$ C\ten{0}{00}=0,~~~~~~C\ten{i}{00}=C\ten{0}{0i}=C\ten{0}{i0}
 =-\oot g^*\_{00,i}=\fs\_{,i},~~~~~~C\ten{i}{0j}=C\ten{i}{j0}=\oot(e\^i\der{j}-e\^j\der{i}), ~~~~~C\ten{0}{ij}=-\oot(e\_{i,j}+e\_{j,i})$$
  \beq  C\ten{i}{jk}=
 \oot (g^*\_{ij,k}+g^*\_{ik,j}-g^*\_{jk,i})=
 \oot (\hs\_{ij,k}+\hs\_{ik,j}-\hs\_{jk,i})
 +\fs\_{,i}\d\_{jk}-\fs\_{,j}\d\_{ik}-\fs\_{,k}\d\_{ij}.
  \eeqno{litara}
  Hence
 \beq \bar C\^0=-\div\ve,~~~~\bar C\^i=\hs\_{ij,j}-\oot\hs\der{i},~~~~C\_0=0,~~~~C\_i= \oot \hs\der{i}-2\fs\der{i} \eeqno{kuitaraw}
 ($\hs=\sum\_{k=1}\^3 \hs\_{kk}$ is the spatial trace of $\hsij$).
All the quadratic scalars, with all possible contractions, reduce in the  NR limit to linear combinations of the following five basic NR scalars:
\beq \bar S_1=-2\gfss+\frac{1}{4}\hs\_{kj,i}(2\hs\_{ki,j}-\hs\_{kj,i})-2\fs\der{i}(\hs\_{ik,k}-\oot \hs\der{i})+\oot(\nabla\times \ve\^*)^2,\eeqno{s1}
\beq \bar S_2=\oot \hs\der{i}(\hs\_{ik,k}-\oot \hs\der{i})-2\fs\der{i}(\hs\_{ik,k}-\oot \hs\der{i}),\eeqno{s2}
\beq \bar S_3=(\hs\_{ik,k}-\oot \hs\der{i})(\hs\_{il,l}-\oot \hs\der{i})-(\nabla\cdot\ve\^*)^2,\eeqno{s3}
\beq \bar S_4=4\gfss+\frac{1}{4}\hs\der{i}\hs\der{i}-2\fs\der{i}\hs\der{i},\eeqno{s4}
\beq \bar S_5=10\gfss+\frac{1}{4}\hs\_{kj,i}(3\hs\_{kj,i}-2\hs\_{ki,j})+2\fs\der{i}(\hs\_{ik,k}-\frac{3}{2}\hs\der{i})-2e\^*\_{i,j}e\^*\_{j,i}.\eeqno{s5}
\par
To get a theory with ``correct'' lensing, $\f$ must be all that remains in the NR limit, the NR solutions of the difference field equations (\ref{surtino}) must thus have $\hs\_{\m\n}=0$, resulting in $\hmn=\hat h\_{\m\n}=0$.
I now proceed to determine the subclass of BIMOND theories for which this is the case.
\par
The NR field equations can be derived directly from the NR Lagrangian
\beq \bar L=-\frac{1}{\epg}\int d\^3x[\gfs+(\grad\fh)\^2+ \azs\M(\bar\Z_1,\bar\Z_2,...)-2\gfss-\f\r-\fh\rh], \eeqno{shamush}
where the variables $\bar\Z_i$ are linear combinations of the five NR basic scalars in Eqs. (\ref{s1})-(\ref{s5}) (in units of $\azs$).
\par
This Lagrangian is obtained from the relativistic one by adhering to the NR approximation. For example, we take the volume element in the interaction term (not in the Einstein terms) to be $1$ because the $\psi$ corrections to it would act like a density of order $\azs\sim\r\_{\Lambda}$.
\par
The schematic form of all the variable $\bar\Z_i$ is
\beq \bar S_m=a\_m\gfss+2b\_m\fs\der{i}(\partial\hs)\_i+c\_m(\partial\hs)^2+d\_m(\partial e\^*)^2, \eeqno{layuta}
with $a\_m,~b\_m,~c\_m,~d\_m$ numerical coefficients.
\par
The $\m\n$ component of the general field equations is gotten by variation of the action over $\gmn$ and $\hgmn$. In the NR case, the $ij$ components are thus gotten by varying over $\hij$  and $\hat\hij$, and the $0i$ components by variation over $e\_i$ and $\hat e\_i$. However, $\f$ and $\fh$ appear in all the diagonal elements as $\f\d\_{\m\n}$; so, varying over $\f$ and $\fh$ gives the sum of the $\m\m$ components of the field equations (not the trace).
\par
Starting with $\ve$, we see from Eq. (\ref{layuta}) that, since $\ve$ does not couple to the other degrees of freedom in any of the NR scalars, the three $0i$ components of the difference equation are of the form
\beq   [\partial(\sum_m  d\_m\pd{\bar\M}{\bar S_m}\partial e\^*)]\_i=0, \eeqno{masesear}
with $\bar\M(\bar S_m)$ the reduction of the interaction function to the NR case and is a function of the NR scalar variables $\bar S_m$.
(There are three indices implicit in Eq. (\ref{masesear}), two from the derivatives and one from $e\^*$; two are contracted and one, $i$, remains free.) Thus, with zero boundary conditions at infinity, this admits the solution $\ve\^*=0$.
\par
This is a general result that applies to any choice of scalars constructed from any power of $C\ten{\l}{\m\n}$. To see why, note in Eq. (\ref{litara}) that $\ve\^*$ appears only in $C\ten{\l}{\m\n}$ with one time component, and that the other nonvanishing components have an even number of time components. Since we are contracting with $\emn$, $\partial\ve\^*$ cannot appear in just one $C\ten{\l}{\m\n}$ factor, lest this scalar involves an odd number of time indices, which cannot be contracted with $\emn$. But then, the $0i$ equations will always be at least linear in $\partial\ve\^*$, and $\ve\^*=0$ obtains. This result is related to the fact that in the NR limit the sources are taken as static, so the matter energy-momentum tensor is invariant to time reversal, under which $\ve\^*$ changes sign.
\par
Variation over $\f,~\fh$ gives a difference equation of the schematic form
\beq   \partial_i\{\sum_m\pd{\bar\M}{\bar S_m}[a\_m\partial_i\fs+b\_m(\partial \hs)\_i]\}\propto \r-\rh,  \eeqno{matushar}
with $(\partial \hs)\_i$ standing for terms with derivatives of $\hs$ with a free index $i$ (such as $\hs\_{ji,j}$ or $\hs\der{i}$).
Varying over $\hij$ and $\hsij$ give the $ij$ components of the difference equation in the schematic form
\beq   \{\partial[\sum_m\pd{\bar\M}{\bar S_m}(b\_m\partial\fs+c\_m\partial \hs)]\}_{ij}=0, \eeqno{matulam}
where the subscript $ij$ can come from derivatives of $\f$ or $h$, or from subscripts of $h$.
\par
Equation (\ref{matulam}) tells us that the obstacle to a good NR theory are the terms with $\partial\fs$, which come from scalars where $\fs$ and $\hsij$ couple: If these terms do not vanish identically, then $\hsij=0$ is not a solution, because it would imply $\partial\fs=0$.
\par
{\it A good NR limit is thus gotten if (and only if)\footnote{But see correction to this statement in the ``Note added'' at the end of the paper.} the NR interaction Lagrangian can be written as a function of  scalars that do not contain such mixed terms.} Equation (\ref{matulam}) is then satisfied for $\hsij=0$, when we impose this as a boundary condition at infinity, for which all $\bar S_m$ in Eq. (\ref{matushar}) becomes $\propto \gfss$, and it becomes a MOND equation sourced by $\r-\rh$
(see details below).
\par
There are two mixed terms that can (and do) appear in the NR scalars (\ref{s1})-(\ref{s5}): $\fs\der{i}\hs\der{i}$ and  $\fs\der{i}\hs\_{ik,k}$. Of the five independent NR scalars we can thus form three independent ones that do not have mixed terms.
The most general such combinations are of the form
\beq S_{s,q,p}=s S_1-q S_2+(q-s)( S_4- S_5)+p S_3,\eeqno{scalar}
with $s$, $q$, and $p$ numbers. The NR limit of such a scalar is
\beq \bar S_{s,q,p}=(4s-6q)\gfss+ (\frac{s}{2} -\frac{3q}{4})\hs\_{ij,k}\hs\_{ij,k}+\frac{q}{2}\hs\_{ij,k}\hs\_{ik,j}-(p+\frac{q}{2})\hs\der{i}\hs\_{ik,k}+ (\frac{p}{4}+
\frac{q}{2}-\frac{s}{4})\hs\der{i}\hs\der{i}+p\hs\_{ik,k}\hs\_{ij,j}.  \eeqno{bityo}
The single scalar variable used in previous work is gotten with $s=q=1$, $p=0$.
\par
It is convenient to also work with the combinations
\beq u\equiv 4s-6q,~~~~~ v\equiv p+\frac{q}{2},  \eeqno{pqq}
because they appear often, and in terms of which
\beq \bar S_{q,u,v}=u\gfss+ \frac{u}{8}\hs\_{ij,k}\hs\_{ij,k}+\frac{q}{2}\hs\_{ij,k}\hs\_{ik,j}-v\hs\der{i}\hs\_{ik,k}+ \frac{1}{4}(v-\frac{u}{4})\hs\der{i}\hs\der{i}+(v-\frac{q}{2})\hs\_{ik,k}\hs\_{ij,j}.  \eeqno{bituv}
We can write
\beq S_{q,u,v}=qS_q+uS_u+vS_v;~~~~~S_q\equiv\oot(3S_1-2S_2-S_3-S_4+S_5),~~~S_u\equiv\frac{1}{4}(S_1-S_4+S_5),~~~S_v\equiv S_3.   \eeqno{gumsha}
\par
For scalars with $u=0$,  $\gfss$ disappears altogether in the NR limit. So to get a meaningful theory we need that at least one of the scalars has $u\not = 0$ (and that cancellation does not occur otherwise).
\par
Alternatively to starting from the NR Lagrangian, we can calculate $S\ten{\c}{\m\n}$ and substitute in the difference field equation (\ref{surtino}). The only relevant components are $S\ten{i}{\m\n}$, because of the time independence. One finds that for the $S_{q,u,v}$ scalars
$S\ten{i}{0j}$ are linear only in $\partial e$, $S\ten{i}{jk}$ only in $\partial\hs$, causing the $0i$ and $jk$ equations to yield $e\^*\_i=\hs\_{ij}=0$, as desired, while $S\ten{i}{00}\propto \fs\der{i}$.
\par
For solutions of the NR field equations (i.e., those with $\hs\_{\m\n}=0$) all the ``good''  scalars reduce to a single scalar
\beq \bar S_{s,q,p}=u\gfss.  \eeqno{bityoma}
\par
I shall restrict myself, hereafter, to the subclass of BIMOND theories where the interaction Lagrangian is a function of only scalars of this type.
\par
To summarize, if the interaction is a function of scalars of the $S_{q,u,v}$ type, the two metrics are of the single-potential type
\beq g\_{\m\n}=\emn-2\f\d\_{\m\n},~~~~~\hat g\_{\m\n}=\emn-2\fh\d\_{\m\n},  \eeqno{sumararara}
with the two potentials determined by the equations
\beq 2\Delta\f+\div[\m(|\grad\fs|/\az)\grad\fs] -\Delta\fs=\epg\r,  \eeqno{nrf}
\beq 2\Delta\hat\f-\div[\m(|\grad\fs|/\az)\grad\fs] +\Delta\fs=\epg\rh,  \eeqno{nrfehat}
\beq \m=\sum_m u\_m\pd{\M}{\Z_m},  \eeqno{jujuju}
with the variables in $\M(\Z_m)$ given by $\Z_m=S_m/\azs$ ($S_m=S_{q\_m,u\_m,v\_m}$), reducing here to $\Z_m=\bar\Z_m=u\_m\gfss/\azs$  ($u\_m=4s\_m-6q\_m$).
\par
Since all variables become equal up to a numerical factor, we can write $\M(\bar\Z_1,\bar\Z_2,...)=\bar\M[\gfss/\azs]$, a function of a single variable. Then
\beq \m(x)=\frac{d\bar \M(z)}{dz}|\_{z=x\^2}.  \eeqno{hurata}
\par
Or, if we express the Lagrangian density as a function of $S_q,~S_u,~S_v$, then
\beq \m(x)=\frac{\partial\M}{\partial \Z_u}|\_{\Z\_u=x\^2,~\Z\_q=\Z\_v=0}.  \eeqno{hurabar}
\par
Taking the difference of the two field equations yields
\beq \div[\m(|\grad\fs|/\az)\grad\fs] =\fpg(\r-\rh), \eeqno{diffa}
which epitomizes the MOND phenomenology, with $\m(x)$ playing the role of the MOND interpolating function.
\par
The sum of the equations gives
 \beq \Delta\f\^+ =\fpg(\r+\rh),  \eeqno{summa}
where $\f\^+\equiv \f+\hat \f$; so $\f\^+$ equals the Newtonian potential sourced by $\r+\rh$.
\par
In the Newtonian limit, $\m(x\rar\infty)\rar 1$, the two sectors decouple, with each potential satisfying its own Poisson equation.
\par
An important lesson from these results is that studying NR systems informs us in a rather limited way on the dependence of the interaction on the different scalars. We can only learn about the behavior of $\M$ on a limited subspace of the variable space through $\bar\M(z)$.
For example, if $\M$ depends on scalars with $u=0$, we can gain no information on this dependence from the NR behavior.
\subsection{The general weak-field case \label{wflscalars}}
For the single-variable case, the WFL of BIMOND, and gravitational waves, were discussed in some detail in Ref. \cite{milgrom14b}. Much of the discussion there carries {\it mutatis mutandis} to the several-variable case. Here I just briefly describe the form of good  scalars in the WFL.
\par
Write in the general WFL
 \beq g\_{\m\n}=\emn+h\_{\m\n},~~~~
 \hat g\_{\m\n}=\emn+\hat h\_{\m\n},
\eeqno{rutsde}
and treat the theory to lowest order in $h\_{\m\n}$ and $\hat h\_{\m\n}$.
We saw that the theory decouples for the perturbations $h\^{\pm}\_{\m\n}=h\_{\m\n}\pm\hat h\_{\m\n}$, where the $+$ sector is equivalent to GR with all the gauge freedom.\footnote{I call the difference here $h\^{-}\_{\m\n}$ to distinguish it from $\hs\_{\m\n}$ used in the NR limit.}
 For the difference sector, the five basic scalars with all possible contractions, reduce in the WFL to
$$ S_1=-\frac{1}{4}h\^-\_{\n\r,\c}h\^-\_{\n\r,\c}+\oot h\^-\_{\n\r,\c}h\^-\_{\n\c,\r},~~~~~~ S_2=\oot h\^-\der{\a}h\^-\_{\a\m,\m}-\frac{1}{4}h\^-\der{\a}h\^-\der{\a},~~~~~~ S_3=h\^-\_{\a\m,\m}h\^-\_{\a\n,\n}+\frac{1}{4}h\^-\der{\a}h\^-\der{\a}-h\^-\der{\a}h\^-\_{\a\n,\n}$$
\beq S_4=\frac{1}{4}h\^-\der{\a}h\^-\der{\a}, ~~~~~~ S_5=\frac{3}{4}h\^-\_{\n\r,\c}h\^-\_{\n\r,\c}-\oot h\^-\_{\n\r,\c}h\^-\_{\n\c,\r},\eeqno{sw5}
where for each pair of repeated indices one is understood to be raised with $\emn$ and the pair summed over.
So, the WFL of the good scalars $S_{s,q,p}$ is
\beq S_{s,q,p}=(\frac{s}{2} -\frac{3q}{4})h\^-\_{\m\n,\l}h\^-\_{\m\n,\l}+\frac{q}{2}h\^-\_{\m\n,\l}h\^-\_{\m\l,\n}-(p+\frac{q}{2})h\^-\der{\m}h\^-\_{\m\n,\n}+ (\frac{p}{4}+\frac{q}{2}-\frac{s}{4})h\^-\der{\m}h\^-\der{\m}+ph\^-\_{\m\n,\n}h\^-\_{\m\l,\l}  \eeqno{bitrosh}
or
\beq S_{q,u,v}=\frac{u}{8}(h\^-\_{\m\n,\l}h\^-\_{\m\n,\l}-\oot h\^-\der{\m}h\^-\der{\m})+\frac{q}{2}(h\^-\_{\m\n,\l}h\^-\_{\m\l,\n}-h\^-\_{\m\n,\n}h\^-\_{\m\l,\l})
+v(h\^-\_{\m\n,\n}-\oot h\^-\der{\m})^2.  \eeqno{bitroshuv}

It would be interesting to investigate the properties of gravitational waves in the theories that employ these scalars.
For example, there might be some particularly happy choices of $s,~q,~p$ in the context of gravitational waves.
Such studies are beyond the scope of this paper.
\par
One interesting result is that scalars that disappear from the NR theory -- and so the dependence on which cannot be deduced from observations of galactic systems -- are still important in the context of gravitational waves. These are the scalars of the type $q=2s/3$ ($u=0$), whose WFL is
\beq  S_{q,0,v}=v(h\^-\_{\m\n,\n}-\oot h\^-\der{\m})^2+ \frac{q}{2}(h\^-\_{\m\n,\l}h\^-\_{\m\l,\n}-h\^-\_{\m\n,\n}h\^-\_{\m\l,\l})  \eeqno{bitmalua}
If, in addition, $q=0$, we have
\beq S\propto(h\^-\_{\m\n,\n}-\oot h\^-\der{\m})^2,  \eeqno{bitgag}
which vanishes if $h\^-\_{\m\n}$ satisfies the harmonic gauge.  Reference \cite{milgrom14b} discussed the relevance of the harmonic gauge, even here, where we do not have gauge freedom in the difference sector.
\section{Some thoughts on BIMOND cosmology   \label{flrw}}
Given the underlying relativistic theory, such as some version of BIMOND, a cosmology does not follow from first principles. As in GR, construction of a cosmology requires much observational input from our real Universe, concerning, e.g., the material content and its properties, and the initial conditions. In GR cosmology, when observations could not be accommodated unless one invokes dominant contribution from dark matter, it was invoked even if its existence is not suggested by any compelling theory. The introduction of a small cosmological constant (or ``dark energy'') contribution has also been forced by observations, despite its constituting a great headache for theory. As regards initial conditions, it was only observations (e.g. of the thermal CMB) that pointed to a beginning in a singularity (as opposed, e.g., to a steady-state Universe), and we are still groping for a concrete initial scenario (such as various models of inflation) to account for various observations, such as the initial injection of (quantum) fluctuations that have led to the observed structure, and the CMB anisotropies.
\par
In the case of BIMOND, the preceding discussion shows that we still have much freedom in choosing the specific version of the theory. This choice concerns whether the theory is symmetric in the two metrics, to the choice of scalar arguments, and to the exact form of the interaction Lagrangian function. And, the need for observational input in constructing a cosmological model is even more acute than in GR, mainly because we have to deal with two sectors, one of which is not even observable at present: Is the material content of the twin sector similar to that in ours? Is it distributed in the same way? Such questions arise naturally\footnote{It was shown in Ref. \cite{milgrom10b}, that in the MOND regime, matter and twin matter repel each other. So structures of the two are expected to be separated. Twin matter also repels (matter) light in the MOND regime; so twin-matter bodies constitute diverging gravitational lenses. We may be able to exploit this to learn about the distribution of twin matter.}
\par
Exploring cosmology within the extended BIMOND framework is far beyond the scope of this paper. This class of theories must admit a large number of cosmological solutions that might be of observational interest.
\par
To get a glimpse of how, for example, the different scalars might enter cosmology in different ways (in the NR limit they all reduce to the same scalar), I consider briefly a class of model solutions. In these, the two metrics are very small departures from some FLRW geometry.
\subsection{Small departures from FLRW}
We seek solutions of extended BIMOND for which there exists a frame in which
\beq \gmn=\rgmn +\hmn,~~~~ \hgmn=\rgmn+\hat h\_{\m\n},  \eeqno{cosmfl}
where $\rgmn$ is of the FLRW form. To simplify the presentation, I specialize to the case where $\rgmn$ has zero spatial curvature:
\beq \rgmn={\rm diag}(-1,a\^2,a\^2,a\^2), \eeqno{cosmara}
with $a(t)$ the cosmological scale factor.
\par
We will consider $h\_{\m\n}$ and $\hat h\_{\m\n}$ -- which are injected as initial conditions of the Hubble-Lemaitre expansion phase from an earlier phase, such as an inflation phase -- to be very small, $h\_{\m\n},~\hat h\_{\m\n}\lll 1$, uncorrelated, random fluctuations whose time and space averages vanish.\footnote{If indeed our cosmology is of this sort, it would justify our taking the background for the weak-field limit to be, locally, a double Minkowski spacetime.}
We also assume that these potentials vary rapidly and have short-wavelengths, namely, that all during the Hubble-Lemaitre expansion phase, the frequencies are much smaller than the expansion rate, and the wavelengths are much shorter than the horizon size: $(\dot a/a)|h\_{\m\n}|\lll |h\_{\m\n,0}|,~a\^{-1}|h\_{\m\n,x}|$ ($x$ are the comoving space coordinates). We will thus neglect everywhere terms of order $h\_{\m\n}$ and $\hat h\_{\m\n}$ themselves, or higher, but not in their derivatives, which are not assumed small.\footnote{If we measure $|\partial h|=\upsilon\az$ in units of $\az$, then, the wavelength $\l\sim |h|/|\partial h|\sim (|h|/\upsilon)\ell\_H$, where $\ell\_H$ is today's Hubble distance, and where I used the coincidence (\ref{coinc}). So $\l\lll\ell\_H$ is equivalent to $|h|\lll\upsilon $.}  These potentials constitute then a background of stochastic gravitational waves, whose propagation properties need to be deduced separately from the cosmological considerations, by studying the WFL on a double Minkowski background (because the wavelength and periods are very small compared with cosmological scales). Because the derivatives of these fluctuations, which enter the interaction Lagrangian, are not small, they can contribute appreciably to the interaction EMT, $\Tmn$, and potentially introduce an extra matterlike contribution, which in turn, affects the evolution of $a(t)$.
\par
Even within this limited class of cosmologies there is a wide scope of possibilities.
I thus further restrict myself to a simple ansatz, where the trial stochastic potentials are of the form
\beq \hmn= -2\t~ {\rm diag}(1,a\^2,a\^2,a\^2),~~~~ \hat h\_{\m\n}= -2\hat\t~ {\rm diag}(1,a\^2,a\^2,a\^2),  \eeqno{cosmfer}
similar to the solutions of the NR limit (but the potentials here are time dependent).
For the ``relative-acceleration tensor''  we then have
\beq C\ten{\l}{\m\n}=\oot\rg\^{\l\s}(\d g\_{\m\s,\n}+\d g\_{\n\s,\m}-\d g\_{\m\n,\s}),  \eeqno{coscon}
where
\beq \d\gmn=\gmn-\hgmn=-2\vrf~ {\rm diag}(1,a\^2,a\^2,a\^2),~~~~~\d\Gmn=\Gmn-\hGmn=2\vrf~ {\rm diag}(1,a\^{-2},a\^{-2},a\^{-2}),   \eeqno{hahay}
and $\vrf=\t-\hat\t$. In expression (\ref{coscon}), terms of order $\vrf\partial\rgmn$ and $\vrf\partial\vrf$ were neglected.
\par
In making this ansatz, I am inspired by the success of Skordis and Zlosnik \cite{sz21} in having propounded, recently, a relativistic version of MOND that achieves several important desiderata. In particular, this theory is said to reproduce the effects of the putative dark matter in cosmology.
In essence, Skordis and Zlosnik \cite{sz21} developed a version of TeVeS using an idea discussed, e.g., in Ref. \cite{scherrer04}, to the effect that a scalar field subject to a certain potential that depends on its gradient can act as cosmological dark matter,'' under certain conditions. Reference \cite{sz21} did not add a new degree of freedom to TeVeS, but rather employed the scalar field that is anyhow needed in TeVeS.
\par
This success brings to mind the possibility that the same idea can be grafted on BIMOND, to reproduce cosmologically viable solutions without new contributions to the matter actions (and energy-momentum tensor), which would count as dark matter. We would want the dark-matter-like contribution to come from the modified gravitational action, namely, to be encapsulated in $\Tmn$.
\par
In BIMOND, we do not have a scalar {\it per se} as a gravitational degree of freedom. But the hope is that there might be cosmologically relevant solutions of BIMOND involving a function that plays the role of the scalar.\footnote{Note that our $\t$, $\hat\t$, and $\vrf$ are not scalars, as the metrics are supposed to have the above form in some specific frame.} So here I am only trying one possible ansatz involving a geometry characterized by a small departure from FLRW, encapsuled in a single function. There may be other such ansatzes.
\par
For the above ansatz, I derive here the field equations for the general choice of scalars, laying the ground for a future more detailed study of cosmological solutions of this type.
\par
For the metrics of the form (\ref{cosmfer}) we have to lowest order in $\vrf$
$$ C\ten{0}{00}=\vrf\der{0},~~~C\ten{0}{0i}=C\ten{0}{i0}=\vrf\der{i},~~~C\ten{0}{ij}=-\dij a\^2\vrf\der{0},$$
\beq C\ten{i}{00}=a^{-2}\vrf\der{i},~~~C\ten{i}{0j}=C\ten{i}{j0}=-\dij\vrf\der{0},~~~
C\ten{i}{jk}=\d\_{jk}\vrf\der{i}-\d\_{ji}\vrf\der{k}-\d\_{ki}\vrf\der{j},
\eeqno{cicici}
and the traces
\beq \bar C\^{0}=-4\vrf\der{0},~~~\bar C\^{i}=0,~~~C\_{0}= -2\vrf\der{0},~~~ C\_{i}=-2\vrf\der{i}. \eeqno{zumzum}
The five basic quadratic scalars (contracted with the reference metric) are then
\beq S_1=-2\rGmn\vrf\der{\m}\vrf\der{\n};~~~~~S_2=8\vrfzs;~~~~~ S_3=-16\vrfzs;~~~~~ S_4=4\rGmn\vrf\der{\m}\vrf\der{\n};~~~~~S_5=10\rGmn\vrf\der{\m}\vrf\der{\n},  \eeqno{lituret}
where, $\rGmn\vrf\der{\m}\vrf\der{\n}= -\vrfzs +\gvrfs$, and $\nabla\vrf$ is the space gradient of $\vrf$ with respect to the proper distances $dl=adx$.
Thus,
\beq S_{s,q,p}=-2(2s+q+8p)\vrfzs +(4s-6q)\gvrfs;~~~~~~~~
 S_{q,u,v}=-(u+16v)\vrfzs +u\gvrfs.  \eeqno{numirafet}
We see that scalars with $u=0$, which disappear from the NR limit, can still be important in cosmology (if $v\not = 0$).\footnote{Taking $a=1$ and $\vrf$ time independent, reduces to the NR case.}
So, we may envisage, for example, a BIMOND version where the interaction depends on two scalars $\M(S_1,S_2)$, with $S_1$ having $u\not=0$, and $S_2$ having $u=0$, with the dependence of $\M$ on them being such that the dependence on $S_2$ enters cosmology, and that on $S_1$ determines the NR limit.
\subsubsection{Field equations}
We need to substitute our ansatz (\ref{cosmfer}) in the field equations and see whether consistent solutions of interest can be gotten.
As before, it is useful to consider separately the sum and difference of the field equations (\ref{melte}) and (\ref{melteha}).
\par
As is done in treating standard cosmology in GR, we average the field equations (designated $\av{}$) over times much shorter than the (instantaneous) Hubble time but much larger than the periods in $\vrf$, and on scales much smaller than the horizon, but much larger than the characteristic wavelengths of $\vrf$.\footnote{There is of course the perennial question -- extensively discussed in the context of GR -- of the legitimacy of replacing, in nonlinear equations, the average of the field equations by the same equations for the averages.}
\par
We consider cosmologies where the two sectors are the same on average (smoothing over the fluctuations). Thus the averages of all quantities are the same in the two sectors, while the fluctuations are uncorrelated.
\par
In the averaged field equations we thus take $\av{T\ten{\m\n}{M}}=\av{\hat T\ten{\m\n}{M}}=\tilde T\ten{\m\n}{M}$.
\par
From expression (\ref{con1}), and since $\av{\Up_s}=-6\av{\vrfzs}-2\av{\gvrfs}$, we have
\beq \av{\hat\T\ten{(1)}{\m\n}}=\av{\T\ten{(1)}{\m\n}}\equiv\tilde\T\ten{(1)}{\m\n}=-\frac{1}{4} [\az^2\av{\M}-6\av{\vrfzs}-2\av{\gvrfs}]\rgmn. \eeqno{shanuva}
\par
As regards the $\Tmn\^{(2)}$ and $\hTmn\^{(2)}$ contributions, we  first calculate $\hat\U\_{\m\n}=\U\_{\m\n}=\tilde\U\_{\m\n}$, which appear in them. Their values for the two sectors are equal because, to our approximation, we use $\rgmn$ for contraction in both.
Substituting from  Eqs. (\ref{cicici}) and (\ref{zumzum}) in expressions  (\ref{mugaga}), the general forms of the $\tilde\U\_{\m\n}$ for the basic scalars $S_m,~~m=1,...,5$, is as follows
\beq \tilde\U\^{(m)}\_{00}=\a\_m\vrfzs+\b\_m \gvrfs;~~~~~~~~~~
\tilde\U\^{(m)}\_{0i}=\z\_m\vrf\der{0}\vrf\der{i};~~~~~~~~~\tilde\U\^{(m)}\_{ij}=\xi\_m\vrf\der{i}\vrf\der{j}+\chi\_m\dij\gvrfs+\o\_ma\^2\dij\vrfzs. \eeqno{gataras}
The coefficient values are given in Table \ref{table1}.
\par
In averaging the $\T\ten{(2)}{\m\n}$ contributions, we note [from Eq. (\ref{lituret})] that the arguments of the derivatives of $\M$ (call them $\M'$ for short here) depend only on $\vrfzs$ and $\gvrfs$, so it is invariant to change of the sign of $\vrf\der{0}$ or of any of the $\vrf\der{i}$. Thus -- since opposite values of these derivatives are equally probable -- upon averaging
$\av{\M'\vrf\der{0}\vrf\der{i}}=0$ and (using also the assumed spatial isotropy) $\av{\M'\vrf\der{i}\vrf\der{j}}=\dij\av{\M'\gvrfs}/3$.
\par
Thus the contributions of the five basic scalars to $\av{\hat\T\ten{(2)}{\m\n}}=\av{\T\ten{(2)}{\m\n}}\equiv \tilde\T\ten{(2)}{\m\n}$ are
 \beq \av{\M'\tilde\U\^{(m)}\_{00}}=\av{\M'(\a\_m\vrfzs+\b\_m \gvrfs)};~~~~~~ \av{\M'\tilde\U\^{(m)}\_{0i}}=0;~~~~~~
 \av{\M'\tilde\U\^{(m)}\_{ij}}=\dij a\^2\{\av{\M'[\o\_m\vrfzs+(\frac{\xi\_m}{3}+\chi\_m)\gvrfs]}\}. \eeqno{gashipoy}
\par
Using the coefficient values in Table \ref{table1}, we then get for the $s,q,p$ scalars
\beq \av{\M'\tilde\U\^{(sqp)}\_{00}}=\av{\M'[(8p+2q-4s)\vrfzs-4q \gvrfs]};~~~~~~ \av{\M'\tilde\U\^{(sqp)}\_{0i}}=0;~~~~~~
 \av{\M'\tilde\U\^{(sqp)}\_{ij}}=\dij a\^2\av{\M'[8p\vrfzs+\frac{2}{3}(q-2s)\gvrfs]}. \eeqno{gasipil}
\par
Thus, $\T\ten{(1)}{\m\n}$, $\T\ten{(2)}{\m\n}$, and the matter EMTs do not contribute to the averaged difference field equation and contribute to the sum equation $2\tilde\T\ten{(1)}{\m\n}$,  $2\tilde\T\ten{(2)}{\m\n}$, and $2\tilde T\ten{\m\n}{M}$, respectively.
Similarly, because $\T\ten{(3)}{\m\n}$ and $\hat\T\ten{(3)}{\m\n}$ are gotten from each other by interchange of the two metrics, their averages are the same, since we assume full symmetry of the averages. Thus $\T\ten{(3)}{\m\n}$ do not contribute to the difference equation.
\par
It can also be shown that the averages of the two Einstein tensors are equal, and equal to that of the reference FLRW metric. Thus, the difference equation is trivially satisfied.
\par
As an interim result, we can write the sum equation as
\beq  \tilde G_{\m\n}+\tilde \T\_{\m\n}+\epg \tilde T\ten{M}{\m\n}=0  ,\eeqno{summuma}
where
\beq \tilde \T\_{\m\n}\equiv \tilde \T\_{\m\n}\^{(1)}+ \tilde \T\_{\m\n}\^{(2)}+ \oot\av{\T\ten{(3)}{\m\n}+\hat\T\ten{(3)}{\m\n}};  \eeqno{gamguga}
so $(\epg)\^{-1}\tilde \T\_{\m\n}$ play the role of a ``phantom matter'' component additional to the matter EMT.
To the above equation we have to add the (averaged) Bianchi identity (see Sec. \ref{bianchiav}).
 \par
We still need to calculate $\av{\T\ten{(3)}{\m\n}+\hat\T\ten{(3)}{\m\n}}$.
From Eq. (\ref{con3}), the main contribution to this sum is of the form
\beq \av{(\M'S\ten{\c}{\m\n})\_{;\c}-(\M'\hat S\ten{\c}{\m\n})\_{:\c}}= \av{[\M'(S\ten{\c}{\m\n}-\hat S\ten{\c}{\m\n})]\der{\c}}+ \av{\M'[S\^{\a}\_{\m\n}(m)C\ten{\c}{\a\c}-S\^{\c}\_{\a\n}(m)C\ten{\a}{\m\c}-S\^{\c}\_{\m\a}(m)C\ten{\a}{\c\n}]}. \eeqno{hahayutsa}
In the first term, $S\ten{\c}{\m\n}$ and $\hat S\ten{\c}{\m\n}$ differ in the contraction metrics appearing in them (the contractions are symmetric in the two metrics). So, to our approximation their difference is of the schematic form $\vrf C\propto\vrf\partial\vrf$. So, the first term is of the schematic form $\av{\partial(\M'\vrf\partial\vrf)}$ which can be seen to vanish in the averaging process.
\par
We thus have
\beq \av{(\M'S\ten{\c}{\m\n})\_{;\c}-(\M'\hat S\ten{\c}{\m\n})\_{:\c}}\equiv 2\av{\M'\tilde\V\ten{(m)}{\m\n}}~~~ {\rm where} ~~~ 2\tilde\V\ten{(m)}{\m\n}\equiv S\^{\a}\_{\m\n}(m)C\ten{\c}{\a\c}-S\^{\c}\_{\a\n}(m)C\ten{\a}{\m\c}-S\^{\c}\_{\m\a}(m)C\ten{\a}{\c\n}, \eeqno{hahayut}
Like the  $\tilde\U\ten{(m)}{\m\n}$, the $\tilde\V\ten{(m)}{\m\n}$ too are quadratic in the $C\ten{\a}{\b\c}$, with contractions employing the reference metric.
They too can be written in the form
\beq \tilde\V\^{(m)}\_{00}=\bar\a\_m\vrfzs+\bar\b\_m \gvrfs;~~~~~~~~~~
 \tilde\V\^{(m)}\_{0i}=\bar\z\_m\vrf\der{0}\vrf\der{i};~~~~~~~~~\tilde\V\^{(m)}\_{ij}=\bar\xi\_m\vrf\der{i}\vrf\der{j}+\bar\chi\_m\dij\gvrfs+\bar\o\_ma\^2\dij\vrfzs. \eeqno{gatamal}
 So,
 \beq \av{\M'\tilde\V\^{(m)}\_{00}}=\av{\M'(\bar\a\_m\vrfzs+\bar\b\_m \gvrfs)};~~~~~~~~~~
 \av{\M'\tilde\V\^{(m)}\_{0i}}=0;~~~~~~~~~
 \av{\M'\tilde\V\^{(m)}\_{ij}}=\dij a\^2\av{\M'[\bar\o\_m\vrfzs+(\frac{\bar\xi\_m}{3}+\bar\chi\_m)\gvrfs]}. \eeqno{gatamaga}
The coefficient values are given in Table \ref{table1}.
\begin{table*}
\footnotesize
\caption{The coefficients for $\U\_{\m\n}$ [Eq. (\ref{gataras})] and for $\V\_{\m\n}$ [Eq. (\ref{gatamal})], and their differences, for the five basic scalars.  \label{table1}}
\begin{tabular}{lcccccccccccccccccc}

\hline
\multicolumn{1}{c} {m}  & $\a\_m$ & $\b\_m$&    $\z\_m$  &  $\xi\_m$ & $\chi\_m$
& $\o\_m$ & $\bar\a\_m$ & $\bar\b\_m$&    $\bar\z\_m$  &  $\bar\xi\_m$ & $\bar\chi\_m$
& $\bar\o\_m$ & $\a'\_m$ & $\b'\_m$&    $\z'\_m$  &  $\xi'\_m$ & $\chi'\_m$
& $\o'\_m$
\\
\hline

   1 &  -4 &  -2  &   -2  &    -6  &  2  &  -2  &  -5 &  -3  &   0  &    -4  &  1  &  -1  & -1 & -1 & 2 & 2 & -1 &  1   \\
   2 &  2 &  2   &  0   &  -4  &  2  &  -2 &  -7/2 &  3  &  3/2   & -2 &  1  &  -1   & -11/2 & 1 & 3/2 & 2 & -1 &  1  \\
   3 &  8 &  0   &  -8  &   0  &  0  &  8 &  4 & 0   &  -4 &  0  & 0  &  4  & -4 & 0 & 4 & 0 & 0 &  -4   \\
   4 &  -4 &  0  &  -4  &  -4  &  0 &  0  &  4 &  -4 &  0  &  0 &  0 &  0  & 8 & -4 & 4 & 4 & 0 &    0   \\
   5 &  -4 &  2  &  -14  &   -2  &  -2  &  2  &  1 &  7  &  -5/2 &   2 &  -1  &  1  & 5 & 5 & 23/2 & 4  & 1 & -1   \\

 \hline
\end{tabular}
\end{table*}

What enter $\T\_{\m\n}$ are $\tilde\W\_{\m\n}\equiv -\tilde\U\_{\m\n}+\tilde\V\_{\m\n}$, whose coefficients are $\a'\_m=\bar \a\_m-\a\_m$, etc. also given in Table \ref{table1}.
\par
Collecting all the terms we have, finally,
\beq \tilde\T\_{\m\n}=\sum_m\av{\frac{\partial \M}{\partial \Z_m}\tilde\W\ten{(m)}{\m\n}}-\frac{1}{4} [\az^2\av{\M}-6\av{\vrfzs}-2\av{\gvrfs}]\rgmn + \av{\Y\_{\m\n}},\eeqno{kukuta}
where
\beq \av{\Y\_{\m\n}}=\oot\av{\Up\_{\m\n}}+\av{\tilde\V\ten{(2)}{\m\n}-\tilde\V\ten{(1)}{\m\n}}.  \eeqno{jutara}
We have
\beq \av{\Up\_{00}}=6\av{\vrfzs}+4\av{\gvrfs};~~~~~~ \av{\Up\_{0i}}=0;~~~~~~
 \av{\Up\_{ij}}=\frac{2}{3}\dij \av{\gvrfs}, \eeqno{gadmanas}
 and
\beq  \av{\tilde\V\ten{(2)}{00}-\tilde\V\ten{(1)}{00}}=\frac{3}{2}\av{\vrfzs}+6\av{\gvrfs}   , ~~~~\av{\tilde\V\ten{(2)}{0i}-\tilde\V\ten{(1)}{0i}}=0,~~~~\av{\tilde\V\ten{(2)}{ij}-\tilde\V\ten{(1)}{ij}}=\frac{2}{3}\dij \av{\gvrfs}.  \eeqno{iuop}
So,
\beq \av{\Y\_{00}}=\frac{9}{2}\av{\vrfzs}+8\av{\gvrfs};~~~~~~ \av{\Y\_{0i}}=0;~~~~~~
 \av{\Y\_{ij}}=\dij \av{\gvrfs}, \eeqno{gaderl}
\par
The first term in Eq. (\ref{kukuta}) is where there is the freedom in choosing the scalar variables, and the dependence of the interaction function on them. It is where future investigations of this approach need to concentrate.
\par
The second term, contributed by $\tilde\T\ten{(1)}{\m\n}$, is a dark-energy-like term. If $\partial\vrf\lesssim\az$ and $\av{\M}\sim 1$, it gives a cosmological-constant-like contribution which satisfies the observed relation (\ref{coinc}). All terms without $\M$ in them result from introducing the $\Up_s$ term in the Lagrangian.
\par
The third term is of the same order as the second, it does not involve $\M$, and is not of the cosmological-constant form.
\par
I leave more detailed studies for future work, but only make the following brief comments here.
If the first term is to stand for cosmological dark matter during the Hubble-Lemaitre expansion phase, as in the analysis of Ref. \cite{sz21}, it needs to overwhelmingly dominate the other terms at early times, and it needs to represent a pressureless fluid. The first condition is fulfilled in the analysis of Refs. \cite{scherrer04,sz21} by invoking a potential function (analogous $\M$ here) that has an infinite logarithmic derivative at some value of its argument (here it is enough that it occurs for one of the arguments of $\M$), and working near this point. Once we ensure that this term dominates, the second requirement can result if $\av{\M'\tilde\W\ten{(m)}{ij}}=0$ for the relevant variable.
This can occur in any of several ways: If all the variables in $\M$ are of the $sqp$ type, then, using the coefficient values in Table \ref{table1}, we see that
\beq \av{\M'\tilde\W\^{(sqp)}\_{00}}=\av{\M'[(-4p+\frac{17q}{2}-4s)\vrfzs+(8s-10q) \gvrfs]};~~~~~~
 \av{\M'\tilde\W\^{(sqp)}\_{ij}}=\dij a\^2\av{\M'[-4p\vrfzs+\frac{2}{3}(s-q)\gvrfs]}. \eeqno{gasimala}
(and $\av{\M'\tilde\W\^{(sqp)}\_{0i}}=0$). Then, if  $\vrfzs$ and $\gvrfs$ are independent, we need to have $s=q$, $p=0$ to have $\av{\M'\tilde\W\ten{(sqp)}{ij}}=0$. This implies that the relevant scalar is the original $\Up$ scalar.
\par
If however,  $\vrfzs$ and $\gvrfs$ are not independent, for example, if the $\vrf$ waves have a specific velocity $c/\eta$; so that $\gvrfs=\eta\^2 \vrfzs$,
we have
\beq  \av{\M'\tilde\W\^{(sqp)}\_{ij}}=\dij a\^2\av{\M'[\frac{2}{3}(s-q)\eta\^2-4p]\vrfzs}, \eeqno{gasiraq}
with the relevant argument in $\M'$ being $\vrfzs$.
Then there is, for example, a scalar that does not contribute in the NR limit (i.e., with $s=3q/2$), and $p=q\eta\^2/12$, for which
$\av{\M'\tilde\W\ten{(sqp)}{ij}}=0$, and the ``density'' $\propto \av{\M'(\vrfzs)\tilde\W\^{(sqp)}\_{00}}\propto \av{\M'(\vrfzs)\vrfzs}$.
\par
In principle, we could have a non-$sqp$ scalar that might be useful in cosmology, and have only a small effect on lensing in the NR limit.
\subsubsection{The average Bianchi identity \label{bianchiav}}
Starting from the general Bianchi identity (\ref{bianch}), we can replace the metric determinants by that of the reference metric in our approximation, yielding ${\T\^{\m\n}}\_{;\n}+{\hat\T\^{\m\n}}\_{~~~:\n}=0$.
Write this as
\beq (\T\^{\m\n}+\hat\T\^{\m\n})\der{\n}+(\T\^{\m\a}+{\hat\T}\^{\m\a})\tilde\C\ten{\n}{\a\n}+(\T\^{\a\n}+{\hat\T}\^{\a\n})\tilde\C\ten{\m}{\a\n}
+\T\^{\m\a}\d\C\ten{\n}{\a\n}+\T\^{\a\n}\d\C\ten{\m}{\a\n}+{\hat\T}\^{\m\a}\d\hat\C\ten{\n}{\a\n}+{\hat\T}\^{\a\n}\d\hat\C\ten{\m}{\a\n}=0, \eeqno{biagara}
where $\d\C\ten{\a}{\b\c}=\C\ten{\a}{\b\c}-\tilde\C\ten{\a}{\b\c}$, $\d\hat\C\ten{\a}{\b\c}=\hat\C\ten{\a}{\b\c}-\tilde\C\ten{\a}{\b\c}$
($\tilde\C\ten{\a}{\b\c}$ is the connection of the reference FLRW metric).
\par
Averaging this relation as before over scales much smaller than cosmologically relevant, but much larger than the wavelengths and periods, the terms with $\d\C$ average to zero. The terms of the form $(\T\^{\m\a}+{\hat\T}\^{\m\a})\tilde\C\ten{\n}{\a\n}$ average to
$\av{\T\^{\m\a}+{\hat\T}\^{\m\a}}\tilde\C\ten{\n}{\a\n}=\tilde\T\^{\m\a}\tilde\C\ten{\n}{\a\n}$, and similarly for the third term.
The average of the first term can be replaced by  $\av{\T\^{\m\n}+\hat\T\^{\m\n}}\der{\n}=\tilde\T\^{\m\n}\der{\n}$, where the derivative here applies to the secular, cosmological changes of the average EMTs.\footnote{For a quantity $A$ such as ours, write $\av{A\der{\n}}=\av{A}\der{\n}+\av{(A-\av{A})\der{\n}}=\av{A}\der{\n}$, since $A-\av{A}$ is fast oscillating, and so are its derivatives.}
The spatial derivatives vanish, because, as usual, we assume spatial homogeneity of the averages.
So, in all, we have for the averaged Bianchi identity
\beq \tilde\T\^{\m\n}\_{~~~|\n}=0,  \eeqno{binava}
where ``$|$'' is the covariant derivative with respect to the reference, FLRW metric. The Bianchi identity thus constitutes the standard cosmological conservation of the phantom-matter component represented by $\tilde\T\^{\m\n}$.

-------------------------------------------------------------------------

Note added: After the paper was published (in Physical Reviews D), I  noticed that the range of scalars that are compatible with ``correct'' lensing can be extended even further. In the discussion below Eq. (\ref{matulam}), I stated, mistakenly, that the employment of scalars of the type (\ref{scalar}) -- which do not contain mixed terms -- is not only sufficient (which it is), but also necessary for correct lensing. Reexamining Eq. (\ref{matulam}), it is, however, obvious, that we can also allow scalars that do contain mixed terms, but do not contain $\gfss$, provided the derivatives of $\bar\M$ with respect to them vanishes when these scalars vanish (which happens when $\hsij=0$ for such scalars).
These scalars are of the form
\beq \tilde S=\a S_2+\b S_4+\c S_5+(2\b+5\c)S_1+\eps S_3. \eeqno{mayure}
\par
There is some overlap between the two types of scalars: those in Eq. (\ref{mayure}), with both $\a=-2\c$ and  $\b=-\c$, are also of the type (\ref{scalar}), as they contain only terms with $\hsij$.
\par
Since these scalars vanish everywhere in the NR limit (where $\hsij=0$), they do not enter nonrelativistic phenomenology, but may be important in cosmology, and in the description of gravitational waves.

\end{document}